\documentclass{article}
\usepackage{graphicx}\usepackage{authblk}
\begin{document}
	\author{B. Das \thanks{bishnu8116@gmail.com}, K. B. Goswami \thanks{koushik.kbg@gmail.com} and P. K. Chattopadhyay \thanks{pkc$_{-}76$@rediffmail.com}}
	\affil{IUCAA Centre for Astronomy Research and Development (ICARD), Department of Physics, Coochbehar Panchanan Barma University, Vivekananda Street, District: Coochbehar, \\ Pin: 736101, West Bengal, India}
	
	\title{A comparative study on maximum mass and radius of compact star from Heintzmann geometry and TOV approach }
	\maketitle
	\begin{abstract}
		In this article a class of anisotropic compact star is analysed in Heintzmann geometry. We have introduced the pressure anisotropy parameter ($\alpha$) and solved Einstein field equations to obtain stellar model. We have considered $g_{tt}$ component as proposed by Heintzmann and by solving Einstein field equation, the $g_{rr}$ component is evaluated in presence of pressure anisotropy. It is noted that for isotropic star ($\alpha=0$), the maximum mass lies within the range $1.87-3.04~ M_{\odot}$ for radii ranges between $8-13$ Km. For anisotropic compact stars maximum mass increases with $\alpha$ and lies within the range $1.99-3.23~ M_{\odot}$ for anisotropy parameter $\alpha=0.5$. The physical viability of the model is examined by applying our model to study the properties of few known compact objects. It is noted that all the stability conditions are fulfilled in the proposed model. It is interesting to note that maximum mass calculated from our model and from solving TOV equation are approximately same and also the predicted radius of few newly observed pulsars and companion star of GW events GW 190814 and GW 170817 from our model comply with the estimated value of radius from observation.
	\end{abstract}
	\section{Introduction}
	Study of astrophysical compact objects like White-Dwarf ($henceforth~$WD), Neutron Stars ($henceforth~$ NS) and Black holes ($henceforth~$BH) etc. have become the keen interest of many astrophysicists in last few decades. This field connects different branches of physics such as astrophysics, particle physics, nuclear physics and high energy physics and acts like as a bridge among them. To explore many physical properties of compact objects, we must encounter the properties of matter at very high density regime higher than the density of nuclear matter. Many astrophysical stellar models have been developed to investigate the physical features of compact objects. For the study of properties of dense matter, neutron stars are the most common and well-understood compact objects. These stars entirely are made of neutrons.
	However, it is not possible to explain the observed properties of many compact object like 4U 1820-30, SAX J 1808.4-3658, HER X-1 and PSR J1614-2230 etc. precisely from the stand point of neutron star model available till date. Such stars have both lower mass and radius than that of NS but their compactness (ratio of mass to radius) is higher in comparison to NS. Observed physical properties of such stars may be explained if we consider the presence of strange matter inside such stars.   
	These stars may be classified known as  Strange Stars ($henceforth~$SS) and are composed of Strange Quark Matter ($henceforth~$SQM) \cite{Haensel}. In case of neutron stars, there is a possibility that SQM may exist inside of such stars in self-bound or slightly unbound form \cite{Haensel,Drago}. In quark star model, massive NS might exist in the form of Strange Quark Stars ($henceforth~$SQS) or quark matter with a thin layer envelope of Hadronic matter. Alcock et al. \cite{Alcock} described different possibilities of conversions of neutron star into strange stars which are: {(i)} u d s quark matter via u d quark matter through pressure-induced transformation, {(ii)} through high-energy neutrino sparking and {(iii)} transformation of neutron matter into strange matter at ultra high densities. In strange stars the weak interaction involving the equilibrium of different quark flavors results in the emission of neutrinos from steller environment. Which subsequently cools down the star and this cooling rate is much higher in SS than in NS \cite{Alcock}. However many researchers discussed in  refs. \cite{Page, Weber, Schaab} that in some cases SQM in SQS's may also cool down slowly and owing to this phenomena their surface temperature may not be distinguishable from some slowly cooling NS. According to Witten \cite{Witten}, it is necessary to consider the metastability of ordinary NSs being realistic provided that quark matter is stable. The equation of state ($henceforth~$EoS) of interior matter at extremely high density greater than nuclear matter density is not known precisely. Therefore, it is very difficult to predict exact mass limit of NS. Considering neutrons as an ideal Fermi gas inside the NS, Oppenheimer and Volkoff \cite{JR} determined the maximum mass of NS and it is approximately $0.7~M_{\odot}$. Usually, quark matter is specified by soft EoS and de-confinement of quark matter inside compact stars is the prominent reason behind the lowering of maximum mass of such kind of stars \cite{Srinivasan}. But in the case of bulk quark or quark-cluster matter there could be stiff EoS due to presence of strong coupling interaction between the nucleons. Rhoades and Ruffini \cite{Rhoades and Ruffini} have obtained the maximum mass of a NS which is $3.2M_{\odot}$, considering perfect fluid distribution interior of NS having matter density higher than the nuclear matter density. Nauenberg and Chapline \cite{Nounberg and Chapline} have predicted that the maximum mass of NS could be as high as of $3.6 M_{\odot}$. A wide range of ($1.46-2.48 M_{\odot}$) of maximum mass of compact stars has been predicted by Haensel et al. \cite{Haensel1} considering different EoS. Maximum mass $2.58~M_{\odot}$ of WD has been obtained by Das et al. \cite{UD} considering magnetic field due to one Landau energy level and continuous accretion of matter from surrounding on outer surface of WD.
	
	In case of highly dense matter inside a compact object, anisotropy may be  developed as proposed by Ruderman \cite{Ruderman1} and Canuto \cite{Canuto}. anisotropic behaviour in pressure may also be developed inside a compact object due to super-conductivity and super-fluidity as indicated by Bowers and Liang \cite{Bowers}. The possible origin of development of pressure anisotropy at such high density regime has been proposed in details by Herrera and Santos in their review work \cite{Herrera1}. Pressure anisotropy may also be originated apart from the reasons due to the following phenomena: (i) pion condensation \cite{Sawyer}, (ii) phase transition \cite{Sokolov} and (iii) presence of type 3A super-fluid \cite{Kippenhahn} or a presence of solid core.
	
	Several anisotropic stellar model has been developed by many researchers and it is established that many physical aspects of stellar configuration found to be compatible with  observational results in presence of pressure anisotropy. In this context, solutions of Einstein Field Equation \\ ($henceforth~EFE$) of compact object has been obtained by Deb et al. \cite{Deb} considering Mak and Harko density profile and pressure anisotropy in the background of MIT bag model EoS. The solution obtained by Deb et al. \cite{Deb} is free from any singularity and the results derived therein are compatible with the data extracted from observations of few compact objects. Considering anisotropic nature of fluid sphere, stellar model have been developed by Marteens \cite{Marteens} and Gokhroo \cite{Gokhroo} admitting uniform and non-uniform density profile. Physical properties of a class of compact objects have been discussed by Thirukkanesh and Maharaj \cite{Thirukkanesh1}, Sharma and Maharaj \cite{Maharaj3} considering linear form of equation of state of interior matter along with anisotropic distribution of pressure. 
	
	Previously, Kalam et al. \cite{kalam} have obtained model of isotropic compact stars using Heintzmann metric \cite{Heintz}. Recently, Goswami et al. \cite{goswami1} and Das et al. \cite{bdas2} have obtained maximum mass and radius of anisotropic strange star using MIT bag model linear EoS. In this article, we present a class of anisotropic compact stars in Heintzmann metric. Here we consider the $g_{tt}$ component of metric potential as proposed by Heintzmann \cite{Heintz} and physically acceptable value of anisotropy $\Delta$ (difference between $p_{t}$ and $p_{r}$). Then we have solved the Einstein Field Equation to determine the value of $g_{rr}$ component in presence of anisotropy. The exact solution is developed which have isotropic counterpart when $\Delta=0$. Using the value of $g_{rr}$ and $g_{tt}$, we have studied different physical properties of compact objects. We have determined the maximum mass of a compact star for arbitrary chosen radius. Two different methods are employed  to determine maximum mass. At first the maximum mass is evaluated considering radial sound velocity $v_{r}^2 \simeq 1$ inside the star for a given choice of radius. In this article, we have presented one interesting result. First EoS is determined using the data set of energy density ($\rho$) and radial pressure ($p_{r}$) for the parameter space used to set up the model. Using this EoS we have solved TOV equation to plot mass-radius relation and to find maximum mass and corresponding radius. The maximum mass derived by using $v_{r}^2 = 1$ and from solving TOV equation are found to be nearly equal.
	
	The paper is  organized as the following : In Sec. 2, The solutions of Einstein field equations is derived for anisotropic fluid. The exact solutions of Einstein Field Equations in Heintzmann geometry and the relevant physical parameters in association with matching conditions are studied in Sec. 3. Physical acceptability of the present model is discussed in Sec.  4.  Physical bounds on the model is discussed in Sec. 5. In Sec. 6, Physical analysis and applications of our model are presented.   In Sec. 7, we have discussed the stability of the model. Tidal Love number and tidal de-formability of compact stars are discussed in Sec. 8. In Sec. 9, a brief discussion of our main findings is given.
	
	\section{Anisotropic compact stellar model and solutions of Einstein field equation}
	
	We consider a static and spherically symmetric metric element for cold compact star defined by 
	\begin{equation}
		ds^2=-e^{\mu}dt^2 + e^{\lambda}dr^2 +r^2 (d\theta^2+ sin^2\theta d\phi^2),\label{eq1}
	\end{equation}   
	here $\mu$ and $\lambda$ are known as metric functions to be determined from boundary conditions.
	
	The energy momentum tensor ($T^i_{j}$) is given by
	\begin{equation}
		T^i_{j}=v^iv_{j}(p_{t}+\rho c^2)-p_{t}\delta^i_{j}-(p_{t}-p_{r})\xi^i\xi_{j},\label{eq2}
	\end{equation}
	where $p_{r}$, $p_{t}$ are radial and transverse pressures respectively and $\rho$ is the energy density, $v^{i}$ is the velocity vector, $\xi^{i}$ is the unit vector in radial direction and $\delta^i_{j}$ is known as Kronecker delta.
	
	The Einstein field equation connecting the matter sector and associated geometry is given as:
	\begin{equation}
		R^i_{j}-\frac{1}{2} R \delta^{i}_{j}=\frac{8\pi G }{c^4} T^{i}_{j},\label{eq3}
	\end{equation}
	here $T^i_{j}$, $R^i_{j}$ and $R$ are the energy momentum tensor, Ricci tensor and Ricci scalar associated with anisotropic fluid respectively. Using eqs.~(\ref{eq1}) and (\ref{eq2}) in eq.~(\ref{eq3}), the EFE reduces to the following set of equations:

	\begin{equation}
		\rho =\frac{\lambda^ \prime e^{-\lambda}}{r} + \frac{1-e^{-\lambda}}{r^2},\label{eq4}
	\end{equation}

	\begin{equation}
		p_{r}=\frac{\mu^ \prime e^{-\lambda}}{r} - \frac{1-e^{-\lambda}}{r^2},\label{eq5}
	\end{equation}

	\begin{equation}
		p_{t}= e^{-\lambda}\left\{\frac{ \mu^{\prime \prime}}{2}+\frac{{\mu^{\prime}}^2}{4}-\frac{\mu^{\prime} \lambda^{\prime}}{4}+ \frac{(\mu^{\prime}-\lambda^{\prime})}{2 r}\right\}.\label{eq6}
	\end{equation}
	Here, we have considered $8\pi G=1$ and $c=1$.
	Now using eqs.~(\ref{eq5}) and (\ref{eq6}) the anisotropic factor $\Delta=(p_{t}-p_{r})$ is given by,
	
	\begin{equation}
		\Delta= e^{-\lambda} \left\{\frac{ \mu^{\prime \prime}}{2}+\frac{{\mu^{\prime}}^2}{4}-\frac{\mu^{\prime} \lambda^{\prime}}{4}- \frac{(\mu^{\prime}+\lambda^{\prime})}{2 r} + \frac{e^{\lambda}-1}{r^2} \right\}.\label{eq7}
	\end{equation}

	\section{Generating exact solution with matching conditions}
	For generating a new class of solution of Einstein field equations, we choose the  $g_{tt}$ component of metric potential proposed by  Heintzmann \cite{Heintz} and consider the form of anisotropic factor $\Delta$ as given below:
	
	\begin{equation}
		e^{\mu(r)}=A^{2}(1+ ax)^{3}\label{eq8}
	\end{equation}
	and
	\begin{equation}
		\Delta(r)=\frac{\alpha a^{2} x}{(1+a x)^2},\label{eq8.1}
	\end{equation}
	where $x=r^2$, $\alpha$ is the anisotropy parameter and $a$ is a constant having  dimension $lenghth^{-2}$. The choice of the form of anisotropy is reasonable in such way that anisotropy factor ($\Delta$) is regular through out the star and also at center $(r=0), ~\Delta=0$ is ensured.
	Now solving eq. ({\ref{eq7}), using eqs. ({\ref{eq8}) and ({\ref{eq8.1}), we have obtained the anisotropic extension of Heintzmann isotropic solution for $g_{rr}$ component in four dimensions as given below:
				\begin{equation}
					e^{-\lambda}=1-\frac{3ax}{2}\left\{ \frac{1+\Lambda(1+4ax)^{-\frac{1}{2}}+\frac{\alpha}{3}}{1+ax}\right\},\label{eq9}
				\end{equation}
				$A$, $\Lambda$ are two dimensionless constants. When $\alpha=0$, the solution reduces to that obtained by Heintzmann \cite{Heintz} for isotropic compact star. These parameters ($A, a$ and $\Lambda$) are used to investigate various physical features of compact objects in presence of anisotropy and also can be used to construct stellar model for known mass of stars considering boundary conditions. On plugging eqs. ({\ref{eq8}) - ({\ref{eq9}) in eqs. ({\ref{eq4}) - ({\ref{eq6}), we get the following expression for energy density, radial pressure, tangential pressure and are given below:
								\begin{equation}
									\rho(x)= \frac{a\left\{g_{1}(x)(1+4ax)^{3/2}+9 \Lambda (1+3ax)\right\}}{2(1+ax)^2(1+4ax)^{3/2}},\label{eq10}
								\end{equation}
								\begin{equation}
									p_{r}(x)=\frac{a \left\{9+\alpha+(7\alpha-9)ax-3\Lambda(1+7ax)(1+4ax)^{-1/2}\right\}}{2(1+ax)},\label{eq11}
								\end{equation}
								\begin{equation}
									p_{t}(x)=p_{r}(x) +\Delta, \label{eq12}
								\end{equation}
								where $g_{1}(x)=4a^{2} x^{2}(3-\alpha)+9-3\alpha+(39-13 \alpha )ax$.
								Anisotropy parameter $\alpha$ is chosen as a free parameter to make our model physically realistic. The arbitrary constants $a$, $A$ and $\Lambda$ can be determined from the matching conditions at the stellar surface as discussed below: 
								\begin{enumerate}
									\item At the surface $(r=b)$ interior solution should be matched with exterior Schwarzschild solution \cite{KS}. In four dimensions (4D), Schwarzschild exterior metric is given by, 
									\begin{equation}	
										ds^{2} = -(1-\frac{2M}{r})dt^{2}+(1-\frac{2M}{r})^{-1}dr^{2}+r^{2}(d\theta^{2}+ sin^2\theta{d\phi}^{2}), \label{eq14}
									\end{equation}
									where $M$ represent the gravitational mass of the star. Then continuity of internal and external metrics at $r=b$ yields:
									\begin{equation}
										e^{\mu(r=b)}=e^{-\lambda(r=b)}=(1-\frac{2M}{b}).\label{eq15}
									\end{equation} 
									Using eqs. (\ref{eq8}) and (\ref{eq9}), we get
									\begin{equation}
										A^{2}\left\{1+ax_{b}\right\}^{3}=1-\frac{3ax_{b}}{2}\left\{ \frac{1+\Lambda(1+4ax_{b})^{-\frac{1}{2}}+\frac{\alpha}{3}}{1+ax_{b}}\right\}=1-\frac{2M}{b}, \label{eq16}
									\end{equation}
									where $x_{b}=b^2$.
									\item At the surface $r=b$, radial pressure ($p_{r}$) which is a monotonically decreasing function of r, vanishes i.e.
									\begin{equation}
										p_{r}(r=b)=0\label{eq17}
									\end{equation}
								\end{enumerate}
								
								\section{Physical acceptibility of the model}
								The following conditions should be satified for well behaved solution of the EFEs,
								\begin{enumerate}
									\item The metric potentials $e^{\lambda}$, $e^{\mu}$ and all the physical parameters such as $\rho$, $p_{r}$, $p_{t}$, $\Delta$ should be well behaved, regular throughout the star.
									\item Pressure ($p_{r}$ and $p_{t}$) and density ($\rho$) should be positive and monotonically decreasing functions of $r$ upto surface of the star i.e. $p_{r}\geq0$,  $p_{t}\geq0$, $\rho\geq0$, $(\frac{dp_{r}}{dr}) < 0$,   $(\frac{dp_{t}}{dr}) < 0$ and $(\frac{d\rho}{dr}) < 0$ . The value of radial pressure ($p_{r}$) must be vanished at the boundary of the star ($r=b$) but tangential pressure ($p_{t}$) need not and may pick up some non-zero value at the boundary of the star which in-turn gives the non-zero value of anisotropy at the surface. At the center of a star $p_{r}$ and $p_{t}$ pick up same value which implies that anisotropy vanishes at the center.
									\item Central pressures $p_{r}(r=0)$,  $p_{t}(r=0)$ and central density $\rho_{c}(0)$ must be finite and positive to make it clear that solutions should be free from any singularities.
									\item For anisotropic compact objects all the energy conditions should be satisfied at all interior points and surface of the star.
									
									\item The causality conditions i.e. the square of the sound velocities such as radial ($v_{r}^{2}$) and tangential ($v_{t}^{2}$) velocities should be less than that of light and must satisfied the following relation $0 \leq v_{r}^{2}=\frac{dp_{r}}{d\rho} \leq 1$ and $0 \leq v_{t}^{2}=\frac{dp_{t}}{d\rho} \leq 1$ throughout the star.
									\item The value of adiabatic index is $\Gamma=\frac{\rho+ p_{r}}{p_{r}} (\frac{dp_{r}}{d\rho})$ of the interior matter for isotropic realistic stellar model.
									The value of adiabatic index ($\Gamma$) of anisotropic matter should be $\Gamma\geq \Gamma^{\prime}_{max}$, where $\Gamma^{\prime}_{max}= \frac{4}{3}-\frac{4}{3}\bigg(\frac{p_{r}-p_{t}}{|p^{\prime}_{r}|r}\bigg)_{max}$ as proposed by Chan et al. \cite{Chan} and $\Gamma^{\prime}_{max}$ should be $\geq \frac{4}{3}$ (Newtonian limit.).
									\item For stable anisotropic matter the Abreu inequality \cite{Abreu} in view of Herrera's cracking condition $ 0 \leq |v_{t}^2-v_{r}^2|\leq 1$, must be satisfied \cite{Herrera}.
									
								\end{enumerate}

								\section{Physical bounds on the model}
								We now discuss the physical features of the interior matter and other physical conditions to check the physical viability of the model. To check the regularity of the gravitational potentials at the centre, we find that $(e^{\mu})_{r=0}=A^2$, $(e^{\lambda})_{r=0}=1$ and $(e^{\mu})^{\prime}_{r=0}=0$, $(e^{\lambda})^{\prime}_{r=0}=0$. Therefore, gravitational potentials and first derivatives of them are regular at the centre. The expression of central density ($\rho_{0}$) and pressure ($p_{0}$) are 
								\begin{equation}
									\rho_{0}=\frac{1}{2}a\left\{9(1+\Lambda)-3\alpha\right\}\label{eq18}
								\end{equation}
								
								\begin{equation}
									p_{0}=\frac{1}{2}a(9-3 \Lambda+\alpha)\label{eq19}
								\end{equation}
								For positive central density $\rho_{0} > 0$, the condition $a>0$ and $\Lambda>(\frac{\alpha}{3}-1)$ are to be satisfied. On the other-hand positivity of central pressure ensures the conditions $a>0$ and $\Lambda<3+\frac{\alpha}{3}$. Therefore, at the center of the star the conditions $\rho_{0} > 0$ and $p_{0} >0$ to be hold simultaneously, we must have $a>0$ and $(\frac{\alpha}{3}-1)<\Lambda<3+\frac{\alpha}{3}$. Zeldovich's condition \cite{Zeldovich1,Zeldovich2} $(\frac{p_{r}}{\rho})_{0}\leq1$ puts an another bound on $\Lambda$ which is $\Lambda \geq \frac{\alpha}{3}$.  Therefore, the range of the parameter $\Lambda$ would be $\frac{\alpha}{3} \leq \Lambda <3+\frac{\alpha}{3}$ for any $\alpha$ so that our model is physically viable.
								Now the expressions for square of radial ($v_{r}^2$) and tangential ($v_{r}^2$) sound velocities are obtained as,
								\begin{equation}
									\resizebox{0.9\textwidth}{!}{$
										v_{r}^2=\frac{dp_{r}}{d\rho}=\frac{(1+4ax)\left\{(1+4ax)^{1/2}f_{1}(x)\right\}+9\Lambda\left\{1-ax(1+14ax)\right\}}{(1+4ax)^{1/2}\left\{axf_{2}(x)+15(\alpha-1)\right\}-9\Lambda\left\{5+ax(30ax+23)\right\}}$}
									\label{eq20a}
								\end{equation}
								\begin{equation}
									\resizebox{0.9\textwidth}{!}{$
										v_{t}^2=\frac{dp_{t}}{d\rho}=\frac{(1+4ax)\left\{(1+4ax)^{1/2}f_{3}(x)\right\}+9\Lambda\left\{1-ax(1+14ax)\right\}}{(1+4ax)^{1/2}\left\{axf_{2}(x)+15(\alpha-1)\right\}-9\Lambda\left\{5+ax(30ax+23)\right\}}$}
									\label{eq20b}
								\end{equation}
								where $f_{1}(x)=27-5\alpha+ax(99-13\alpha)-7a^{2}x^{2}(4\alpha+9)$, \\ $f_{2}(x)=16a^{2}x^{2}(\alpha-3)+ax(47\alpha-264)-123$ and 
								$f_{3}(x)=27-7\alpha +ax (99-19 \alpha)-36 a^2 x^2(1-\alpha)$.
								
								\begin{table*}[ht!]
									\caption{ Estimated values of $\Lambda$, $A$, $a~(Km^{-2})$, maximum mass $M(M_{\odot})$, compactness $(u_{max})$ and surface red-shift $(Z_{s})_{max}$ for different values of $\alpha~(=0.0, 0.3, 0.5)$ and radius ($b$) of the compact stars  $8$,  $9$, $10$, $11$ and $13$ Km respectively.}\label{Tab1}
									\begin{centering}
										\begin{tabular}{|c|c|c|c|c|c|c|c|} \hline
											\multicolumn{1}{|c|}{Radius } & \multicolumn{1}{|c|}{anisotropy } & \multicolumn{1}{|c|}{$\Lambda$}& \multicolumn{1}{|c|}{$A$}& \multicolumn{1}{|c|}{$a$}& \multicolumn{1}{|c|}{$M_{max}$} & \multicolumn{1}{|c|}{$u_{max}$} & \multicolumn{1}{|c|}{$Z_{s,max}$}  \\
											$b~(Km)$ & ($\alpha$)&  & & $(Km^{-2})$  & $(M_{\odot})$ &   & \\ \cline{1-8} \hline
											& $0.0$ & 0.447535 & 0.279759  &  0.00913  & $1.87$                & $0.3443$     &      $0.7925$    \\ 
											8	& $0.3$ & 0.470361 & 0.289166 & 0.01126   & $1.94$                & $0.3576$     &      $0.8742$    \\ 
											& $0.5$ & 0.488146 &   0.304064 &  0.01323  & $1.99$                & $0.3667$     &      $0.9367$    \\ \cline{2-8}
											\hline
											& $0.0$ & 0.447535 &0.279759 & 0.00722   & $2.10$                & $0.3443$     &      $0.7925$    \\ 
											9	        & $0.3$ & 0.470361&0.289166 &  0.00889 & $2.18$                & $0.3576$     &  $0.8742$    \\
											
											& $0.5$   & 0.488146 & 0.304064  &0.01045  & $2.23$                & $0.3667$     &      $0.9367$    \\ \cline{2-8}
											\hline
											& $0.0$  & 0.447535 & 0.279759 & 0.00584  & $2.33$                & $0.3443$     &      $0.7925$    \\ 
											10	& $0.3$ & 0.470361&0.289166 & 0.00720   & $2.43$                & $0.3576$     &      $0.8742$    \\
											& $0.5$   &0.488146 & 0.304064  & 0.00847 & $2.49$                & $0.3667$     &      $0.9367$    \\ \cline{2-8}
											\hline
											& $0.0$  & 0.447535 & 0.279759 & 0.00483  & $2.57$               & $0.3443$     &      $0.7925$   \\ 
											11	            & $0.3$ & 0.470361& 0.289166 & 0.00595  & $2.67$                & $0.3576$     &      $0.8742$ \\
											
											& $0.5$   &0.488146 & 0.304064  &0.00699 & $2.73$                & $0.3667$     &      $0.9367$    \\ \cline{2-8}
											\hline
											& $0.0$  & 0.447535 & 0.279759 & 0.00346  & $3.04$                & $0.3443$     &      $0.7919$    \\ 
											13	      & $0.3$ & 0.470361& 0.289166 & 0.00426  & $3.15$                & $0.3576$     &      $0.8742$  \\
											
											& $0.5$ & 0.488146& 0.304064  & 0.00501 & $3.23$                & $0.3667$     &      $0.9367$    \\ \cline{2-8}									\hline		    		    		
										\end{tabular}
									\end{centering}
								\end{table*}
								
								\begin{figure}[ht!]
									\begin{center}
										\includegraphics[width=8.3cm]{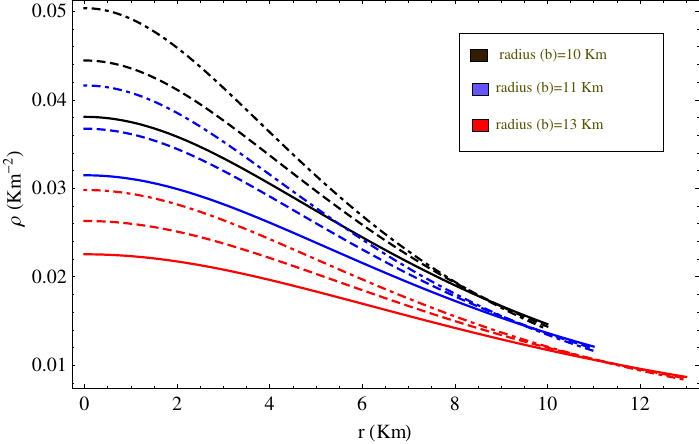}
										\caption{Radial variations of energy density $(\rho)$. Solid, dashed and dot-dashed lines correspond to (i)$~\alpha=0.0$, $\Lambda=0.447535$, (ii) $\alpha=0.3$, $\Lambda=0.470361$ and (iii) $\alpha=0.5$, $\Lambda=0.488146$ respectively and are given in Table-\ref{Tab1}.}
										\label{density}
									\end{center} 
								\end{figure}
								
								\begin{figure}[ht!]
									\begin{center}
										\includegraphics[width=8.3cm]{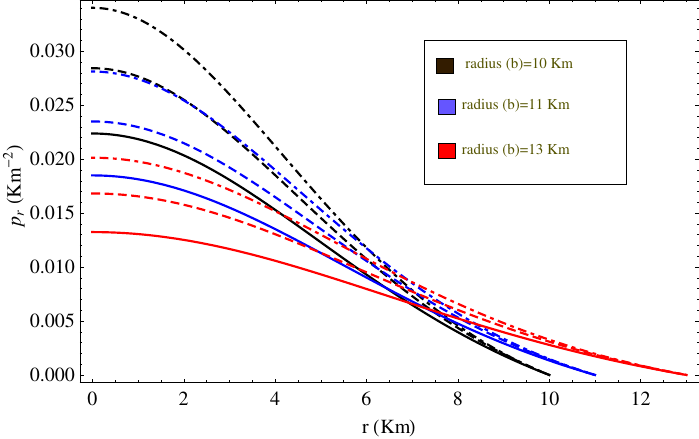}
										\caption{Radial variations of radial pressure $(p_{r})$. Solid, dashed and dot-dashed lines correspond to (i)$~\alpha=0.0$, $\Lambda=0.447535$, (ii) $\alpha=0.3$, $\Lambda=0.470361$ and (iii) $\alpha=0.5$, $\Lambda=0.488146$ respectively and are given in Table-\ref{Tab1}.}
										\label{pr}
									\end{center} 
								\end{figure}

								\begin{figure}[ht!]
									\begin{center}
										\includegraphics[width=8.3cm]{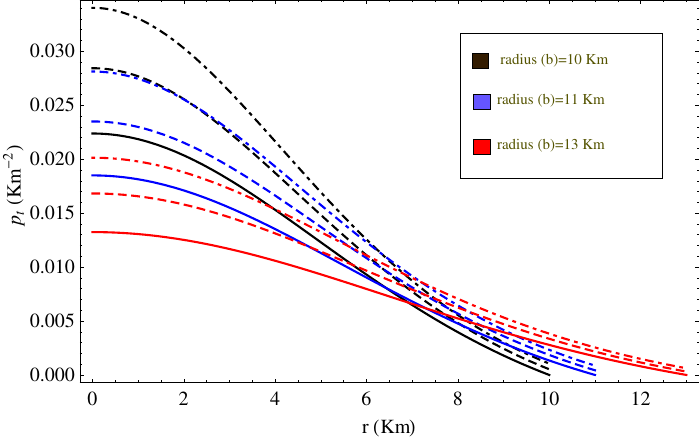}
										\caption{Radial variations of tangential pressure $(p_{t})$. Solid, dashed and dot-dashed lines correspond to (i)$~\alpha=0.0$, $\Lambda=0.447535$, (ii) $\alpha=0.3$, $\Lambda=0.470361$ and (iii) $\alpha=0.5$, $\Lambda=0.488146$ respectively and are given in Table-\ref{Tab1}.}
										\label{pt}
									\end{center} 
								\end{figure}
								
								\begin{figure}[ht!]
									\begin{center}
										\includegraphics[width=8.3cm]{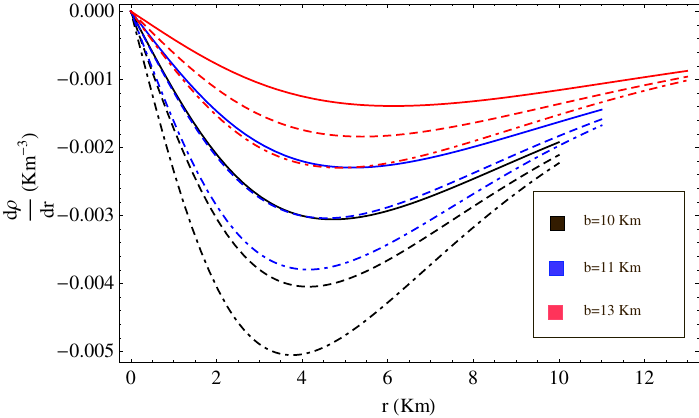}
										\caption{Radial variations of energy-density gradient $(\frac{d\rho_{r}}{dr})$. Solid, dashed and dot-dashed lines represent respectively for (i)$~\alpha=0.0$, $\Lambda = 0.447535$, (ii) $\alpha=0.3$, $\Lambda=0.470361$ and (iii) $\alpha=0.5$, $\Lambda=0.488146$ are given in Table-\ref{Tab1}.}
										\label{densitygradiant}
									\end{center} 
								\end{figure}
								
								\begin{figure}[ht!]
									\begin{center}
										\includegraphics[width=8.3cm]{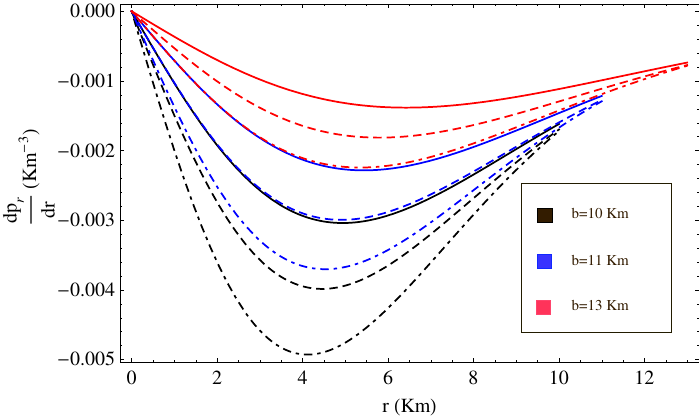}
										\caption{Radial variations of pressure gradient $(\frac{dp_{r}}{dr})$. Solid, dashed and dot-dashed lines corresponds to (i)$~\alpha=0.0$, $\Lambda = 0.447535$, (ii) $\alpha=0.3$, $\Lambda=0.470361$ and (iii) $\alpha=0.5$, $\Lambda=0.488146$ respectively and are given in Table-\ref{Tab1}.}
										\label{prgradiant}
									\end{center} 
								\end{figure}
								
								\begin{figure}[ht!]
									\begin{center}
										\includegraphics[width=8.3cm]{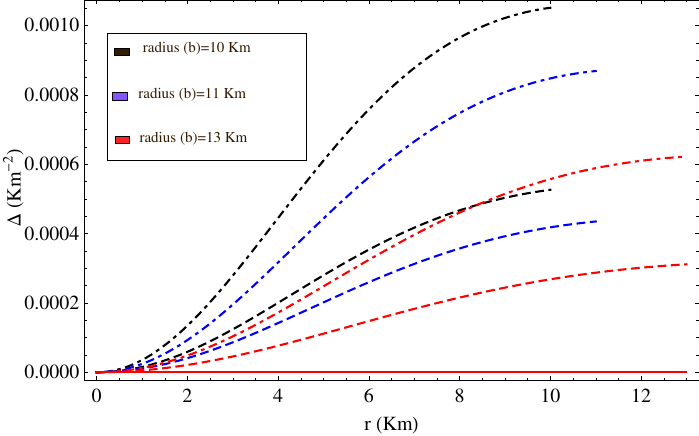}
										\caption{Radial variations of anisotropy $(\Delta)$. Solid, dashed and dot-dashed lines correspond to $(i)~\alpha=0.0$, $\Lambda$=0.447535, (ii) $\alpha=0.3$, $\Lambda = 0.470361$ and (iii) $\alpha=0.5$, $\Lambda=0.488146$ respectively and are given in Table-\ref{Tab1}.}
										\label{delta}
									\end{center} 
								\end{figure}

								\section{Physical analysis and applications}
								In this model, we have developed the stellar configuration of anisotropic compact objects in modified form of Heintzmann geometry \cite{Heintz}. For physical analysis, we have studied various physical parameters namely energy density $(\rho)$, radial and tangential pressures ($p_{r}, p_{t}$), anisotropy ($\Delta$), mass ($M$), compactness ($u=\frac{M}{b}$), surface red-shift ($Z_{s}$) etc. for a compact star.

								The variations of $\rho$, $p_{r},~ p_{t},~\frac{d\rho_{r}}{dr},~\frac{dp_{t}}{dr}$ and $\Delta$ have been shown in Figs. \ref{density} - \ref{delta} for a class of compact stars of different radius ($b=10,11, 13~Km$). All the plots have been shown for the choice of constants ($a, \Lambda, A$ and $\alpha$) for which square of radial sound velocity represented as $v_{r}^2$ is maximum ($\simeq 1$) at the centre or any internal point of the star. To evaluate the maximum mass ($M_{max}$) of a compact star having radius $b$, we first calculate the value of constant $\Lambda$ for which square of sound velocity $v_{r}^2=\frac{dp_{r}}{d\rho}$ is maximum at the centre or any interior point. It is evident from eqs. ({\ref{eq20a}) and ({\ref{eq20b}) that as $v_{r}^2 > v_{t}^2$ always, so it is necessary and sufficient to consider the value of $v_{r}^2=1$ for the determination of maximum mass. We have determined the maximum mass of compact objects for different values of anisotropy parameter $\alpha$ and are tabulated in Table-\ref{Tab1}.	
										\begin{figure}[ht!]		
											\begin{center}
												\includegraphics[width=8.3cm]{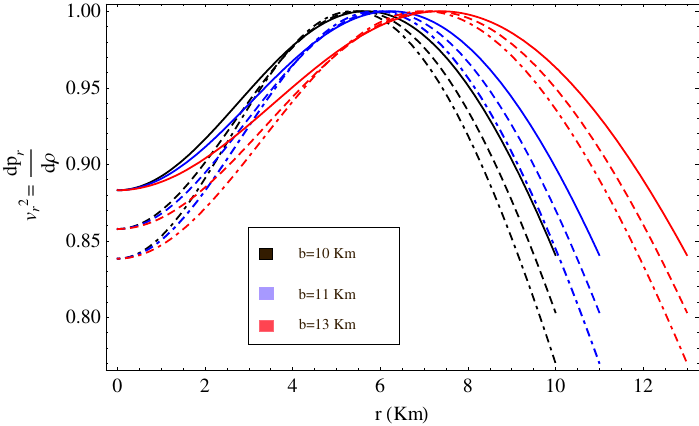}
												\caption{Plot of square of radial sound velocity $v_{r}^2=(\frac{dp_{r}}{d\rho})$. Solid, dashed and dot-dashed lines represent respectively for $(i)~\alpha=0.0$, $\Lambda$=0.447535, (ii) $\alpha=0.3$, $\Lambda=0.470361$ and (iii) $\alpha=0.5$, $\Lambda=0.488146$ are given in Table-\ref{Tab1}.}
												\label{figvr}
											\end{center} 
										\end{figure}
										
										\begin{figure}[ht!]
											\begin{center}
												\includegraphics[width=8.3cm]{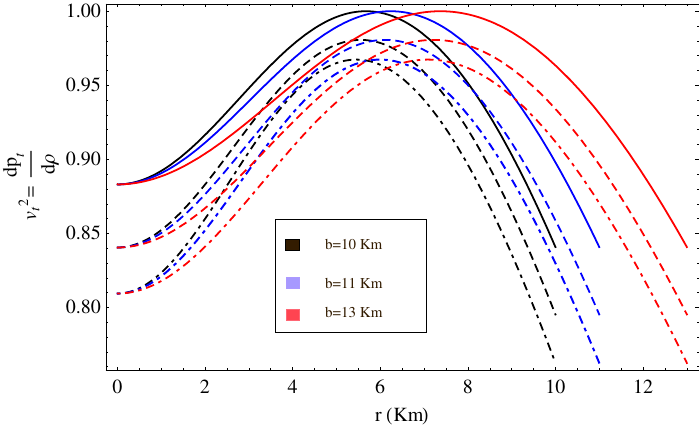}
												\caption{Plot of square of tangential sound velocity $v_{t}^2=(\frac{dp_{t}}{d\rho})$. Solid, dashed and dot-dashed lines correspond to $(i)~\alpha=0.0$, $\Lambda$=0.447535, (ii) $\alpha=0.3$, $\Lambda=0.470361$ and (iii) $\alpha=0.5$, $\Lambda=0.488146$ respectively and are given in Table-\ref{Tab1}.}
												\label{figvt}
											\end{center} 
										\end{figure}
										
										\subsection{Causality condition}
										The sound velocities must be causal throughout the interior of a compact star. The fulfilment of causality conditions is required for a physically viable stellar configuration. For a compact star with pressure anisotropy, two sound velocities namely radial ($v_{r}^2$), tangential ($v_{t}^2$)  should satisfy the conditions $ v_{r}^2=(\frac{dp_{r}}{d\rho})\leq1$ and $ v_{t}^2=(\frac{dp_{t}}{d\rho})\leq1$. In this model, we have plotted the causality conditions in the case corresponding to the maximum mass of compact star and are shown in Figs. \ref{figvr} and \ref{figvt} respectively. 
										
										\begin{figure}[ht!]
											\begin{center}
												\includegraphics[width=8.3cm]{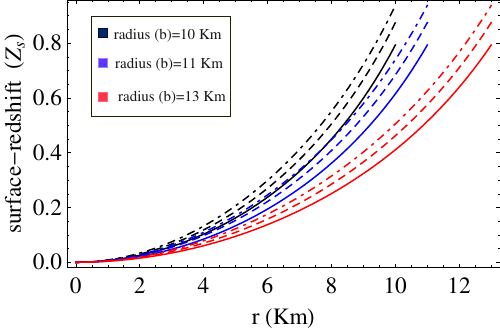}
												\caption{Radial variations of surface red-shift $Z_{s}(r)$. Solid, dashed and dot-dashed lines correspond to $(i)~\alpha=0.0$, $\Lambda$=0.447535, (ii) $\alpha=0.3$, $\Lambda=0.470361$ and (iii) $\alpha=0.5$, $\Lambda=0.488146$ respectively and are given in Table-\ref{Tab1}.}
												\label{figz}
											\end{center} 
										\end{figure}
										
										\subsection{Mass-radius relation and surface red-shift}
										The gravitational mass ($M$) contained within a spherical region of radius $b$ can be
										\begin{eqnarray}
											M(b)&=&4\pi\int_{0}^{b}\rho(r^{\prime}){r^{\prime}}^2dr^{\prime} \nonumber \\
											&=& \frac {abx_{b}\left\{(3-\alpha)(1+4ax_{b})+3\Lambda \sqrt {1+4ax_{b}}\right\}} {4(1+5ax_{b}+4x_{b}^2)},\label{eq21}
										\end{eqnarray}
										where $x_{b}=b^2$. Using Eq. (\ref{eq21}), we get the expression of compactness by the ratio of mass to radius derived as,
										\begin{eqnarray}
											u&=&\frac{M(b)}{b} \nonumber \\
											&=&  \frac {ax_{b}\left\{(3-\alpha)(1+4ax_{b})+3\Lambda \sqrt {1+4ax_{b}}\right\}} {4(1+5ax_{b}+4x_{b}^2)}.\label{eq22}        
										\end{eqnarray}
										Buchdahl\cite{Buchdahl} proposed that for spherically symmetric static fluid sphere maximum allowed value of mass-radius ratio ($u$) is $\frac{4}{9}$. Our model satisfies this condition for the parameter space used here to construct the model.
										The expression for the surface red-shift ($Z_{s}$) of star is given by the following relation where $u$ is taken from eq. (\ref{eq22})
										\begin{equation}
											Z_{s}=(1-2 u)^{-\frac{1}{2}}-1 \label{eq23} 
										\end{equation}
										Surface red-shift ($Z_{s}$) is shown in Fig. \ref{figz}.
										
										\begin{figure}[ht!]
											\begin{center}
												\includegraphics[width=8.3cm]{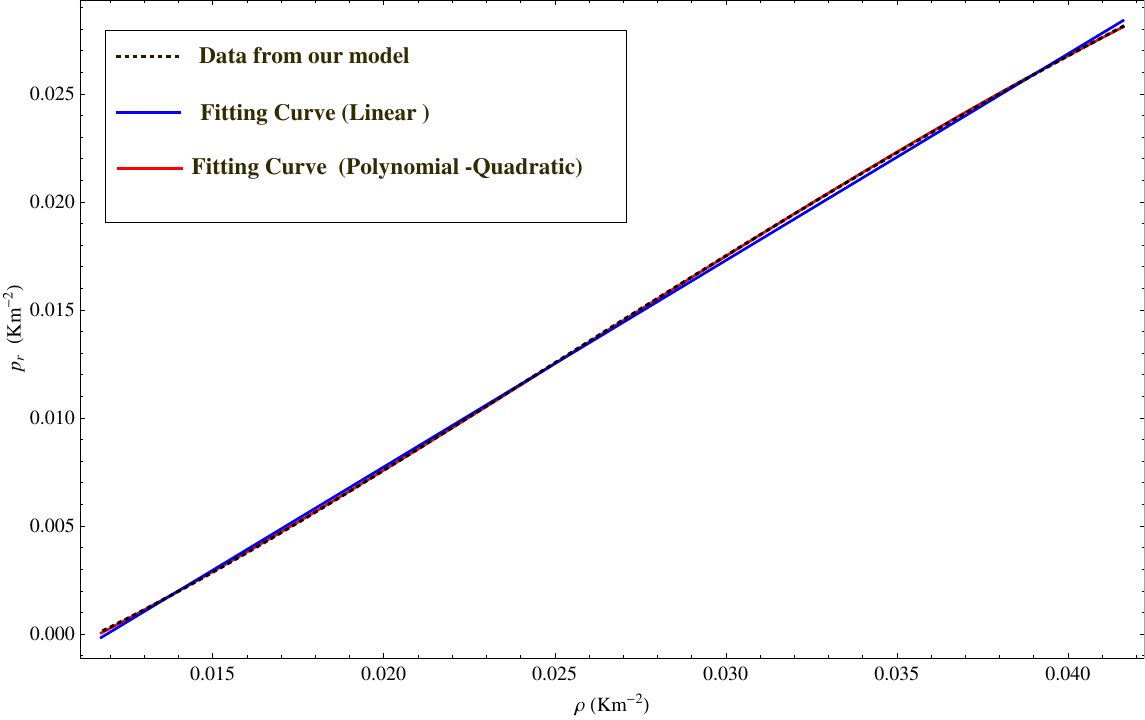}
												\caption{Plot of Equation of state (EoS) inside the compat stars having radius $11$ Km.}\label{figeospolb11}
											\end{center} 
										\end{figure}

										\subsection{Equation of State}
										We have determined the nature of equation of state for interior matter using the data set of $\rho$ and $p_{r}$. Using the method of curve fitting two possible EoSs of interior matter are obtained: one is linear and other is quadratic for arbitrary choice of radius and corresponding maximum mass and are shown in Fig.~\ref{figeospolb11}.  We note that best fitted EoS in our model is quadratic. We have also shown the variation of mass with central density in Figs.~\ref{figmvdtov} and \ref{figmvdtovq}. From Figs.~\ref{figmvdtov} and ~\ref{figmvdtovq}, it is also noted that our model satisfies the Harrison-Zel'dovich-Novikov criterion \cite{Harrison,Zeldovich}. The effects of anisotropy ($\alpha$) on the values of maximum central densities for linear and quadratic EoS are tabulated in Table \ref{Tab5}. 
										
										\begin{table*}[ht!]
											\caption{Estimated values of maximum central densities $(\rho_{c})_{ max}$ for different values of $\alpha$.}\label{Tab5}
											\begin{centering}
												\begin{tabular}{|c|c|c|c|} \hline
													Radius & Anisotropy &	Equation of state & ($\rho_{c, max} \times 10^{15} $)  \\ 
													& & ($EoS$)  &   \\
													& ($\alpha$) &  & ($ gm/cm^{3}$) \\ \cline{1-4}
													& & $p_{r}=-0.01772 +0.96879\rho$ &2.989 \\ 
													&0.0 &$p_{r}=-0.01815 +0.99587\rho-0.40567{\rho}^2$ & 2.955	\\ \cline{2-4}
													& & $p_{r}=-0.01738 +0.96158\rho$ &2.948 \\
													9 & 0.3 &$p_{r}=-0.01795 +0.99592\rho-0.462638{\rho}^2$ & 2.915		\\ \cline{2-4}
													& &$p_{r}=-0.01701 +0.95577\rho$ &2.928 \\
													& 0.5 & $p_{r}=-0.01769 +0.99389\rho-0.47092{\rho}^2$ & 2.914	\\		 \hline		 
													& &$p_{r}=-0.01187+0.96882\rho$ &1.997 \\
													& 0.0 & $p_{r}=-0.01220 +0.99997\rho-0.69043{\rho}^2$ & 1.970	\\ \cline{2-4}
													& & $p_{r}=-0.01164 +0.96162\rho$ &1.977 \\
													11 & 0.3 & $p_{r}=-0.01196 +0.99173\rho-0.61460{\rho}^2$ & 1.949		\\ \cline{2-4}
													& &$p_{r}=-0.011393 +0.95581\rho$ &1.957 \\
													& 0.5 & $p_{r}=-0.01189 +0.99772\rho-0.76680{\rho}^2$ & 1.929	\\		 \hline							    
												\end{tabular}
											\end{centering}
										\end{table*}
										
										\begin{table*}[ht!]
											\caption{ Comparision of Maximum mass determined by solving TOV equation using linear and quadratic EoS with the maximum mass obtained by maximising the sound velocity ($v_{r}^2 = 1$).}\label{Tab4}
											\begin{center}
												\begin{tabular}{|c|c|c|c|c|c|c|} \hline
													Anisotropy &   \multicolumn{2}{|c|} {Maximum mass from our model }	& \multicolumn{4}{|c|}{ Maximum mass from TOV equation}  \\ \cline{4-7}
													&    &  &   \multicolumn{2}{|c|} { Linear EoS} & \multicolumn{2}{|c|} {Quadratic EoS} \\  \cline{2-7}
													$(\alpha)$  & Radius ($b$) &	$M_{max}$  & 	$M_{max}$& Radius ($b$)   & 	$M_{max}$ & Radius ($b$)   \\ 
													&  ($Km$) &	$(M_{\odot})$  &  	$(M_{\odot})$ & ($Km$)  &  $(M_{\odot})$ & ($Km$)  \\ \hline
													
													0.0   &      & 2.10 & 2.12 & 8.89  &   2.14 & 8.94   \\ 
													0.3   &  9    &2.18 & 2.13 &8.93  &   2.15 & 8.99   \\  
													0.5   &      & 2.23 & 2.14 & 8.99  &  2.16 &  9.00   \\ 	\cline {1-7}					    
													
													0.0   &      &  2.57 & 2.59 & 10.95  &  2.62 & 10.94      \\ 
													0.3   & 11    & 2.67 &  2.61& 10.97 &   2.64 & 10.99   \\  
													0.5   &      &  2.73 &  2.62 & 10.99  &   2.65  & 11.07   \\ 	\cline {1-7}						    
												\end{tabular}
											\end{center}
										\end{table*}
										
										\begin{figure}[ht!]
											\begin{center}
												\includegraphics[width=8.3cm]{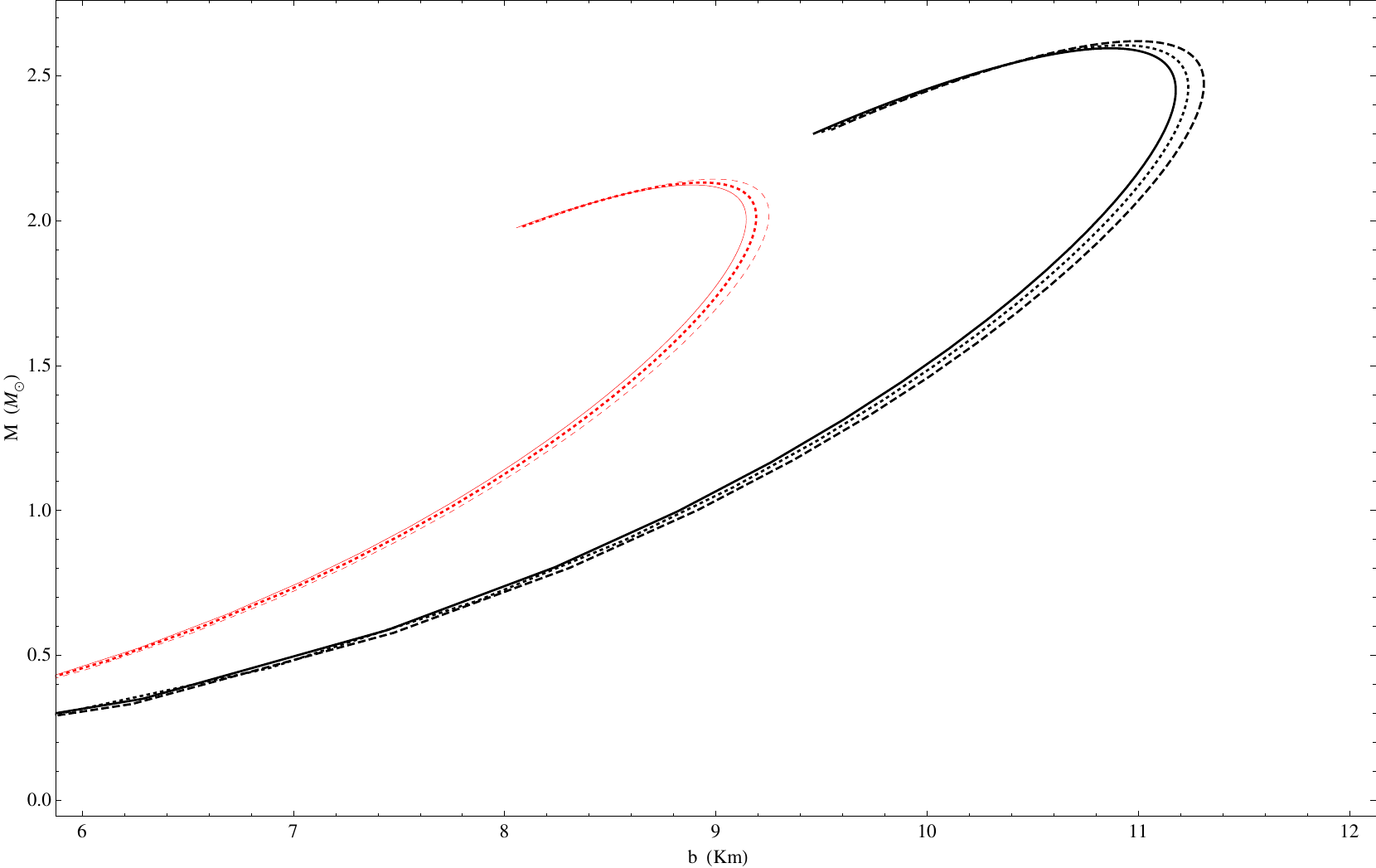}
												\caption{Mass-radius curve for TOV equation. Black and red lines for radius 11 and 9 Km respectively for linear EoS. Solid, dotted and dashed lines for the data when $\alpha=0.0, 0.3, 0.5$ respectively.}\label{figmrtov}
											\end{center} 
										\end{figure}
										
										\begin{figure}[ht!]
											\begin{center}
												\includegraphics[width=8.3cm]{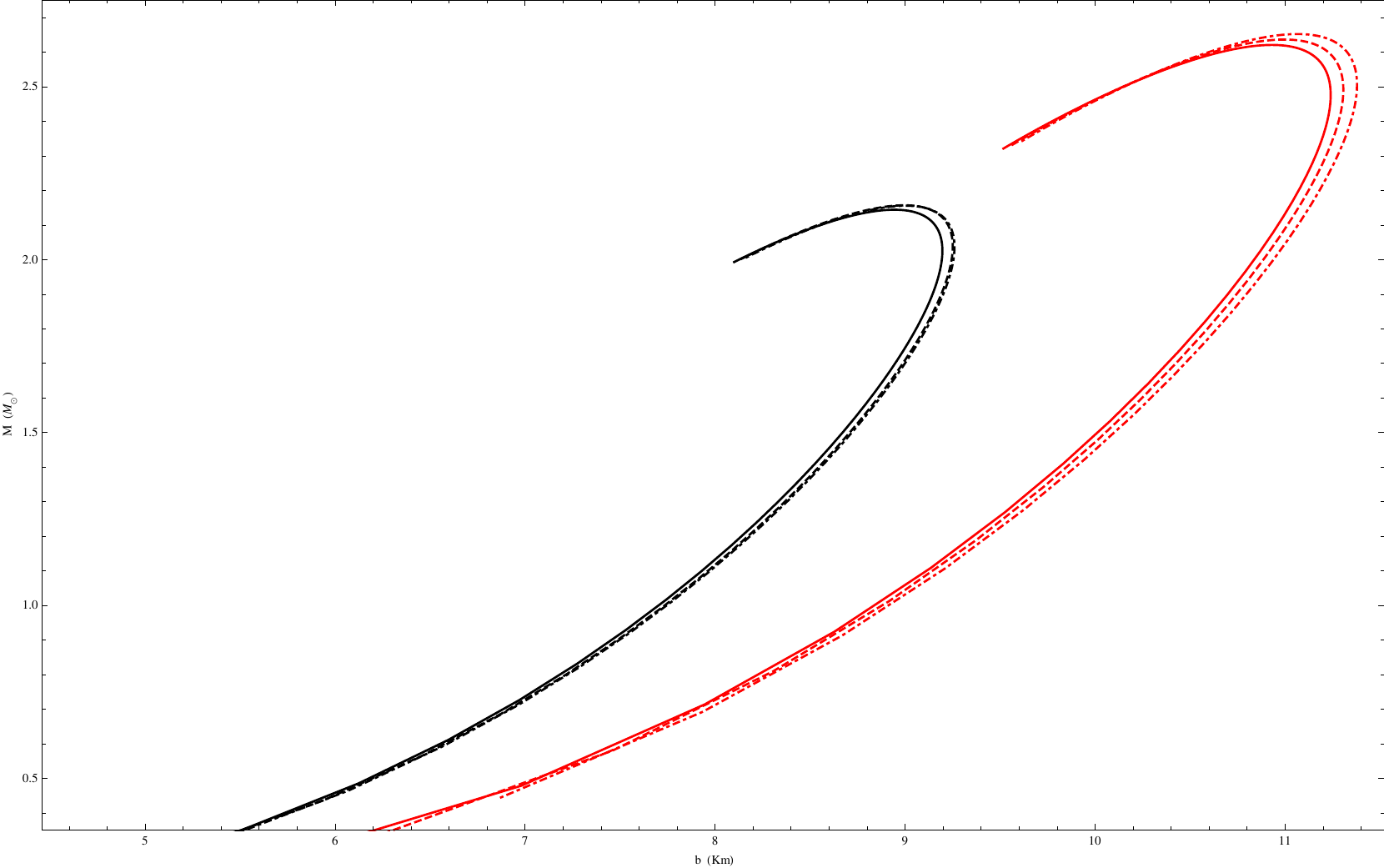}
												\caption{Mass-radius curve for TOV equation. Black and red lines for radius 9 and 11 Km respectively for quadratic EoS. Solid, dotted and dot-dashed lines for the data when $\alpha=0.0, 0.3, 0.5$ respectively.}\label{figmrtovq}
											\end{center} 
										\end{figure}

										\begin{figure}[ht!]
											\begin{center}
												\includegraphics[width=8.3cm]{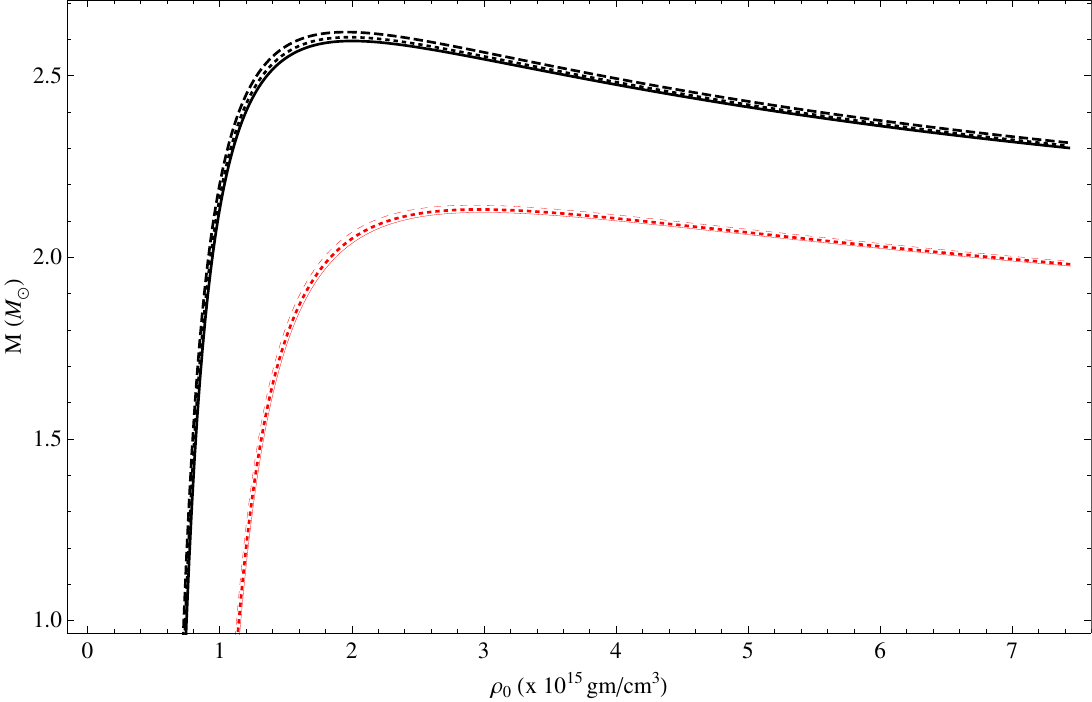}
												\caption{Variation of mass $M (M_{\odot})$ with central density $\rho_{0}$ ($gm/cm^{3}$) in this compact star model. Black and red lines for radii 11 and 9 Km respectively for linear EoS. Solid, dotted and dashed lines for the data when $\alpha=0.0, 0.3, 0.5$ respectively.}\label{figmvdtov}
											\end{center} 
										\end{figure}
										
										\begin{figure}[ht!]
											\begin{center}
												\includegraphics[width=8.3cm]{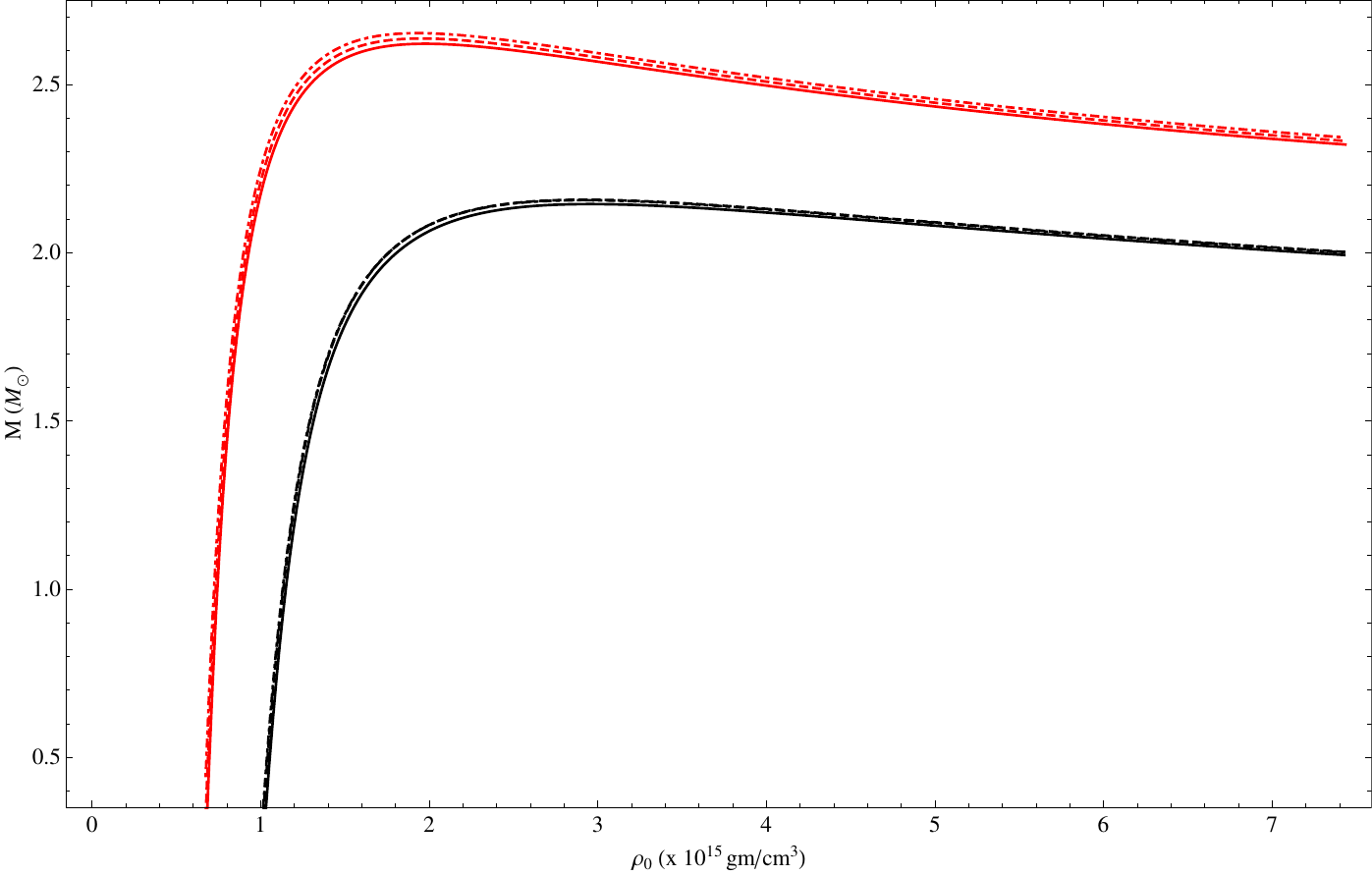}
												\caption{Variation of mass $M (M_{\odot})$ with central density $\rho_{0}$ ($gm/cm^{3}$) in this compact star model. Black and red lines for radii 9 and 11 Km respectively for quadratic EoS. Solid, dotted and dot-dashed lines for the data when $\alpha=0.0, 0.3, 0.5$ respectively.}\label{figmvdtovq}
											\end{center} 
										\end{figure}
										
										\subsection{Maximum mass}
										In this article we have evaluated the maximum allowable mass for the constrained value of parameter space used here by maximising sound velocity ($v_{r}^2=1$) inside the compact objects. Here we consider the radius of the star as input parameter values of maximum mass for different radius are tabulated Table \ref{Tab1}. It is noted that maximum mass increases with anisotropy as expected. After evaluation of maximum mass for chosen radius, we have determined the value of $\rho$ and $p_{r}$ at all interior points and using the values of $\rho$ and $p_{r}$, the EoS is plotted. Linear and quadratic fitted EoS is then used to solve TOV equation to verify the value of maximum mass and radius. Mass-radius curve are shown in Figs.~\ref{figmrtov} and ~\ref{figmrtovq} for linear and quadratic fitted EoS respectively. We have noted some interesting results tabulated in Table \ref{Tab4}. The maximum mass obtained from geometrical consideration by maximising sound velocity is nearly equal to the maximum mass obtained from TOV equation using quadratic fitted EoS for the parameter space used here. Also the maximum radius from TOV equation approaches the arbitrary chosen value for higher anisotropy parameter. 
										
										\begin{figure}[ht!]
											\begin{center}
												\includegraphics[width=8.3cm]{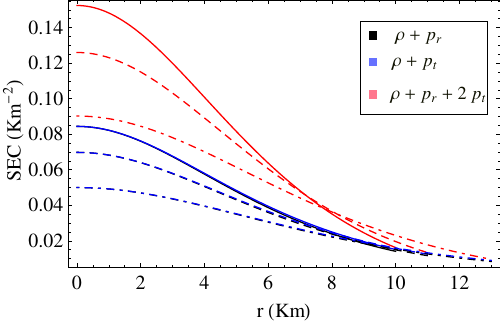}
												\caption{Radial variations of Strong Energy Condition (SEC) in units of $Km^{-2}$ for $\alpha=0.5$ and $\Lambda=0.488146$. Here, solid, dashed and dotdashed lines for radius of the compact star $10$, $11$ and $13$ Km respectively.}\label{figsec}
											\end{center} 
										\end{figure}
										
										\begin{figure}[ht!]
											\begin{center}
												\includegraphics[width=8.3cm]{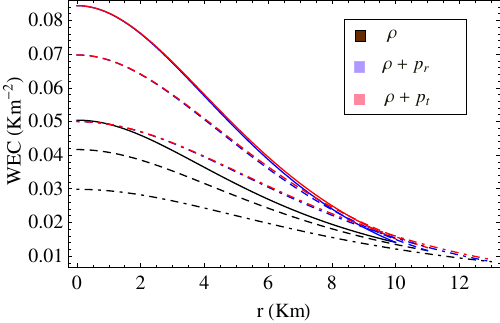}
												\caption{Radial variations of Weak Energy Condition (WEC) in units of $Km^{-2}$ for $\alpha=0.5$ and $\Lambda=0.488146$. Here, solid, dashed and dot-dashed lines for radius of the compact star $10$, $11$ and $13$ Km respectively.}
												\label{figwec}
											\end{center} 
										\end{figure}
										
										\begin{figure}[ht!]
											\begin{center}
												\includegraphics[width=8.3cm]{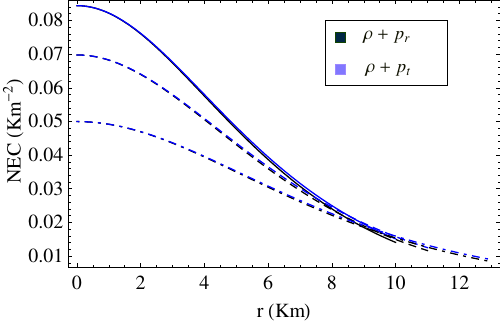}
												\caption{Radial variations of Null Energy Condition (NEC) in units of $Km^{-2}$. for $\alpha=0.5$ and $\Lambda=0.488146$. Here, solid, dashed and dot-dashed lines for radius of the compact star $10$, $11$ and $13$ Km respectively.}\label{fignec}
											\end{center} 
										\end{figure}
										
										\begin{figure}[ht!]
											\begin{center}
												\includegraphics[width=8.3cm]{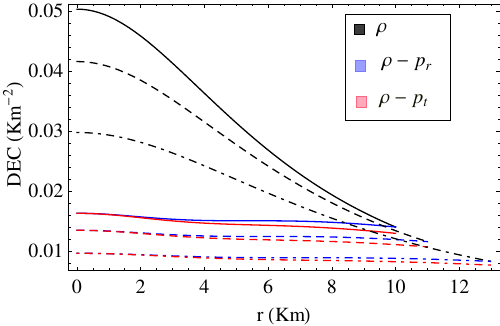}
												\caption{Radial variations of Dominant Energy Condition (DEC) in units of $Km^{-2}$ for $\alpha=0.5$ and $\Lambda=0.488146$. Here, solid, dashed and dot-dashed lines for radius of the compact star $10$, $11$ and $13$ Km respectively.}\label{figdec}
											\end{center} 
										\end{figure}

										\subsection{Energy condition}
										The fulfilment of energy condition is one of the important criterion for any stellar configuration of isotropic or anisotropic matter \cite{Carroll,Pant}. Such energy conditions are Strong Energy Condition or SEC, Weak Energy Condition or WEC, Null Energy Condition or NEC and Dominant Energy Condition or DEC. The mathematical expressions of these energy conditions are expressed as \cite{Brassel1, Brassel2}: 
										
										\begin{enumerate} 
											\item  SEC: $(\rho +p_{r}) \geq 0, (\rho +p_{t}) \geq 0, (\rho +p_{r}+ 2p_{t}) \geq 0$. 
											\item  WEC: $\rho \geq 0, (\rho +p_{r}) \geq 0, (\rho +p_{t}) \geq 0$.
											\item  NEC: $(\rho +p_{r}) \geq 0, (\rho +p_{t}) \geq 0$.
											\item  DEC: $\rho \geq 0, (\rho -p_{r}) \geq 0, (\rho -p_{t}) \geq 0$. 
										\end{enumerate}
										SEC, WEC, NEC and DEC are plotted in Figs. \ref{figsec} - \ref{figdec}. It is observed that all of these energy conditions are satisfied in our model.

										\section{Stability analysis of the model}
										A necessary condition in building stellar model is that a stellar configuration of isotropic/anisotropic fluid should be stable. The stability of such models are studied through the following analysis: 
										\begin{enumerate}
											\item Fulfilment of Tolman-Oppenheimer and Volkoff (TOV) equation
											\item Herrera cracking condition and
											\item Adiabatic index for the fluid.
										\end{enumerate}
										\subsection{TOV Equation}
										The equilibrium between different forces acting on the fluid inside the stellar object should follow the TOV equation. We have used the generalized form of TOV equation as given by Tolman \cite{RC} \& Oppenheimer and Volkoff \cite{JR} and is given below,
										\begin{equation}										-\frac{M_{G}(r)(\rho+p_{r})}{r^2}e^{\frac{(\lambda-\mu)}{2}}-\frac{dp_{r}}{dr}+\frac{2}{r}(p_{t}-p_{r})=0,\label{eq24}
										\end{equation}
										where $M_{g}(r)$ is known as the active gravitational mass  which can be obtained from the Tolman-Whittaker formula \cite{Grøn} and contained within a spherical volume of radius $r$. It is expressed in terms of the metric potential $\mu$ and $\lambda$ and is given below,
										\begin{equation}
											M_{G}(r)=\frac{1}{2}r^{2}\mu^{\prime}e^{\frac{(\mu-\lambda)}{2}}.\label{eq25}
										\end{equation}
										Plugging the value of $M_G(r)$ from eq.~(\ref{eq25}) into eq.~(\ref{eq24}), we have obtained the following relation
										\begin{equation}
											-\frac{\mu^{\prime}(\rho+p_{r})}{2}-\frac{dp_{r}}{dr}+\frac{2}{r}\Delta=0, \label{eq26}
										\end{equation}
										where $\Delta=p_{t}-p_{r}$ and can be obtained from eq.~(\ref{eq7}). Eq.~(\ref{eq25}) has three parts namely (i) the gravitational force $(F_g=\frac{ -\mu^{\prime}(\rho+p_{r})}{2})$, (ii) the hydrostatic force $(F_h= -\frac{dp_{r}}{dr})$ and (iii) the anisotropic force $(F_a=\frac{2\Delta}{r})$.\\
										Therefore, for equilibrium, 
										\begin{equation}
											F_{g}+F_{h}+F_{a}=0.\label{eq27}
										\end{equation}
										\begin{figure}[ht!]
											\begin{center}
												\includegraphics[width=8.3cm]{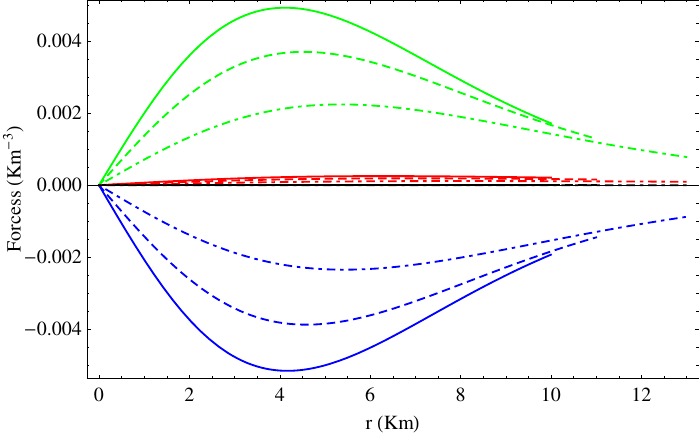}
												\caption{Variations of different forces inside the compact star of different radius for $\alpha=0.5$ and $\Lambda=0.488146$. Solid, dashed and dot-dashed lines for radius $10, 11$ and $13$ Km respectively. Blue, green and red lines for gravitational force $(F_{g})$, hydrostatic force $(F_{h})$ and anisotropic force $(F_{a})$ respectively.}\label{figtov}
											\end{center} 
										\end{figure}
										In Fig.~\ref{figtov}, we have shown the radial variation of $F_{g}$, $F_{h}$ and $F_{a}$ with radial distance. It is noted that the stellar configuration remains in stable equilibrium under the collective effects of force due to gravity, hydrostatic force and force due to anisotropy in pressure at all interior points of a star. From Fig.~\ref{figtov}, it is evident that gravitational force is always larger than hydrostatic force ($F_{h}$) and anisotropic force ($F_{a}$). Also the sum of numerical value of $F_{h}$ and $F_{a}$ is equal to the gravitational force ($F_{g}$). It is also evident that the gravitational force ($F_{g}$) and hydrostatic force ($F_{h}$) increase from centre, attain a maximum value somewhere inside the star and finally decrease up to the surface whereas the anisotropic force ($F_{a}$) first decreases and then increases monotonically  towards the boundary of the star. Such features have previously been predicted by Maurya et al. \cite{Maurya}.
										
										\begin{figure}[ht!]
											\begin{center}
												\includegraphics[width=8.3cm]{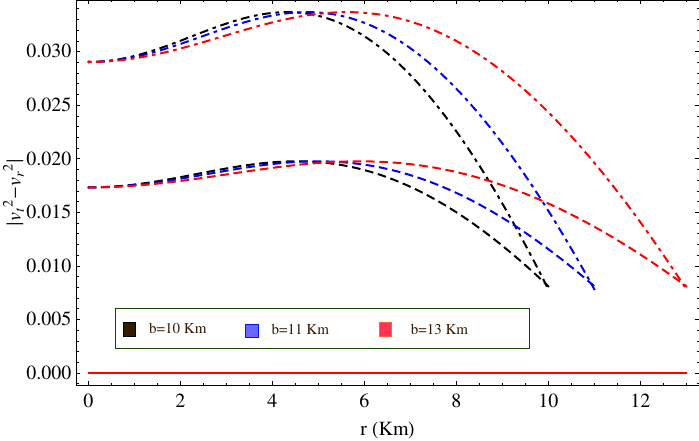}
												\caption{Variations of $|v_{t}^{2}-v_{r}^2|$ inside the compact star of different radius. Plots with solid, dashed and dot-dashed lines for $(i)~\alpha=0.0$, $\Lambda$=0.447535, (ii) $\alpha=0.3$, $\Lambda=0.470361$ and (iii) $\alpha=0.5$, $\Lambda=0.488146$ respectively.}
												\label{fighra}
											\end{center} 
										\end{figure} 
										
										\subsection{Herrera cracking condition}
										Any stellar configuration (isotropic/anisotropic)  should be in stable equilibrium against the fluctuations of its physical quantities and its stability can be checked through Herrera's cracking condition \cite{Herrera}. Based on the Herrera's concept, Abreu et al. \cite{Abreu} introduced a criteria to explain the stability of stellar configuration. Abreu et al. \cite{Abreu} established that a stellar model is found to be stable if the following condition is fulfilled throughout the star
										\begin{equation}
											0 \leq |v_{t}^{2}-v_{r}^2| \leq 1. \label{eq28}
										\end{equation}
										In Fig.~\ref{fighra}, the radial variation of  $|v_{t}^{2}-v_{r}^2|$ has been shown for compact objects having different radius ($b=10,11, 13$ Km). 
										
										\begin{figure}[ht!]
											\begin{center}
												\includegraphics[width=8.3cm]{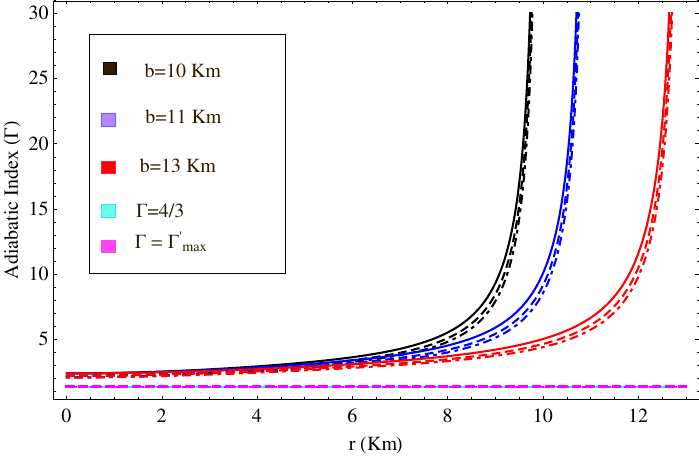}
												\caption{Plot of radial variations of adiabatic index ($\Gamma$) inside the compact stars having radii $10,11$ and $13$ Km. Plots with solid, dashed and dot-dashed lines for $(i)~\alpha=0.0$, $\Lambda$=0.447535, (ii) $\alpha=0.3$, $\Lambda=0.470361$ and (iii) $\alpha=0.5$, $\Lambda=0.488146$ respectively are given in Table-\ref{Tab1}. Here, Dotted and dashed for $\Gamma^{\prime}_{max}$ when $\alpha=0.3$ and $0.5$ respectively.}\label{figadia}
											\end{center} 
										\end{figure}
										
										\subsection{Adiabatic index}
										Adiabatic index ($\Gamma$) is an another important parameter to check the stability of fluid sphere. In case of isotropic star the adiabatic index ($\Gamma$) is expressed as
										\begin{equation}
											\Gamma=\frac{\rho+p_{r}}{p_{r}}\frac{dp_{r}}{d\rho}=\frac{\rho+p_{r}}{p_{r}}v_{r}^{2}.\label{eq29}
										\end{equation}
										According to Heintzmann and Hillebrandt \cite{Heintzmann}, the dynamical value of the adiabatic index ($\Gamma$) is bound to be greater than $\frac{4}{3}$ for any stable and viable stellar model. 
										However, Chan et al. \cite{Chan} outspread the idea of adiabatic index for anisotropic fluid. According to Chan et al.\cite{Chan}, the maximum value of adiabatic index for  anisotropic fluid, where both radial ($p_{r}$) and transverse ($p_{t}$) pressures exist is given below:
										\begin{equation}
											\Gamma^{\prime}_{max}=\frac{4}{3}-\frac{4}{3}\bigg(\frac{p_{r}-p_{t}}{|p^{\prime}_{r}|r}\bigg)_{max} \label{eq59}
										\end{equation}
										In Fig.~\ref{figadia}, the radial variation of adiabatic index ($\Gamma$) for compact stars having radii $10,11$ and $13$ Km has been shown.

										\section{\normalsize Tidal Love numbers and Tidal de-formability}\label{sec8}
										The surface of a compact object might be distorted from its spherical shape under the influence of an external perturbing field due to gravity and this distortion is measured by the parameter $k_2$ known as Tidal Love number. Another parameter known as Tidal de-formability represented by $\chi$ is related to the tidal Love number ($k_2$). The relation between them is given below:
										\begin{equation}
											k_2=\frac{3}{2}\chi\bigg(\frac{M}{b}\bigg)^5.\label{Eq043}
										\end{equation}
										
										Here $M$ is the mass of the deformed compact object and $b$ its radius. Even parity ($l=2~(even~parity)$) Tidal Love number is evaluated by Hinderer et al. \cite{Hinderer1} which has the following form:   
										
										\begin{equation}
											k_2=\frac{8u^5}{5}(1-2u)^2[2+2u(h-1)-h] \times f(u), \label{Eq044}
										\end{equation}
										where 
										$f(u)=\bigg[2u\left\{6-3h+3u(5h-8)\right\}+4u^3\left\{13-11h+u(3h-2)\\ +2u^2(1+h)\right\}+3(1-2u)^2\left\{2-h+2u(h-1)\right\}log(1-2u)\bigg]^{-1}$,\\
										here $u$ is termed as compactness (=$\frac{M}{b}$) and $h=\frac{bH^{'}(b)}{H(b)}$. We have used the expression of $H(r)$ as given in the Ref.\cite{Hinderer2} as follows   
										\begin{eqnarray}H(r)=c_1\big(\frac{r}{M}\big)^2\bigg(1-\frac{2M}{r}\bigg) \times G(r)+ 3c_2 R(r)\label{Eq045}
										\end{eqnarray}
										where $G(r)=-\frac{M(M-r)(2M^2+6Mr-3r^2)}{r^2(2M-r)^2} +\frac{3}{2}log\bigg(\frac{r}{r-2m}\bigg) $ and $R(r)=(\frac{r}{M}\big)^2\bigg(1-\frac{2M}{r}\bigg)$.
										From binary neutron star merger event GW170817 \cite{Abbott1}, Abbott et al. \cite{Abbott2} have predicted some constraints on $\chi$. Bauswein et al. \cite{Bauswein} shown that for the value of $1.4 M_{\odot}$, the value of tidal de-formability should be less than 800. i.e. $\chi<800$. In the Table \ref{Tab3}, the value of tidal Love number and corresponding tidal de-formability of three stars are tabulated. It is observed from Table \ref{Tab3} that the tidal Love number ($k_2$) and the tidal de-formability ($\chi$) pick up lower values when anisotropy parameter ($\alpha$) increases.
										
										\begin{table*}
											\caption{ Estimated values of  Tidal love number and Tidal de-formability in our model.}
											\label{Tab3}
											\begin{center}    
												\begin{tabular}{|c|c|c|c|c|c|} \hline \cline{1-6}
													Compact Stars & Observed & Anisotropy  &  Predicted   & Tidal & Tidal  \\ 
													& mass  &  &  radius ($R_{b}$)       &  Love number  & de-formability  \\ 
													& ($M_\odot$) & ($\alpha$) & $(Km)$ & $(k_2)$ & $(\chi)$ \\ \cline{1-6}   
													
													&   &0.0 & 7.32  & 0.214130 & 79.899    \\     
													{GW 170817}\cite{Abbott1}	&  1.4 &0.3 & 6.90  & 0.163254 & 45.333 \\ 
													&      &0.5 & 6.64  & 0.146273 &  33.521         \\ \cline{1-6}
													&   & 0.0 & 10.3 & 0.207005 & 77.229    \\  
													{PSR J1614-2230}\cite{MC Miller}	          & 1.97 &0.3 & 9.72  & 0.166981& 46.625 \\ 
													&  &0.5 & 6.64  & 0.153284  & 35.251    \\ \cline{1-6}    
													&   & 0.0 &12.28 & 0.197468 & 73.768    \\ 
													{PSR J0952-0607}\cite{RW Romani}       & 2.35&0.3 & 11.59  & 0.178216 & 49.656 \\ 
													&  &0.5 & 11.13  & 0.160038 &  36.417   \\	\cline{1-6}  								
												\end{tabular}
											\end{center}
										\end{table*}
										
										\begin{table*}[ht!]
											\caption{ Estimated value of radii for few compact objects from our model.}
											\label{Tab2}
											\begin{center}
												\begin{tabular}{|c|c|c|c|c|} \hline
													\multicolumn{1}{|c|}{Compact Star } &   {anisotropy } &  \multicolumn{3}{|c|}{$\Lambda=1.0772$} \\ \cline{3-5}
													& ($\alpha$)  & $a$ ($\times 10^{-4}$) & $A$ ($\times 10^{-2}$) & {Predicted}  \\ 
													& & $ (Km^{-2})$ &   &  {radius} (Km) \\ \cline{1-5} 
													& $0.0$     & $25.9293$       & $45.8484$ &  $10.30$    \\ 
													PSR J1614-2230 \cite{MC Miller} & $0.3$     & $34.9669$       & $41.3254$ & 9.72      \\ 
													$(M=1.97 M_{\odot})$  	& $0.5$     & $43.1412$       & $38.0655$ & 9.35    \\ \cline{2-5}
													\hline		
													& 0.0     & 18.2216      & $45.8484$ &  $12.29$    \\ 
													PSR J0952-0607 \cite{RW Romani} & $0.3$     & $24.5728$       & $41.3254$ &11.59     \\ 
													$(M=2.35 M_{\odot})$ 	& $0.5$     & $30.3172$       & $38.0655$ &11.13  \\ \cline{2-5}
													\hline		
													& $0.0$     & $15.0011$       & $45.8484$ &  $13.54$     \\ 
													GW 190814 \cite{Abbott1} & $0.3$     & $20.2298$       & $41.3254$ & $12.77$  \\ 
													$(M=2.59 M_{\odot})$ 	& $0.5$     & $24.9589$       & $38.0655$ & 12.29  \\ \cline{2-5}
													\hline		
													& $0.0$     & $51.3413$       & $45.8484$ &  $7.32$     \\ 
													GW 170817 \cite{Abbott2} & $0.3$     & $68.2363$       & $41.3254$ & 6.90      \\ 
													$(M=1.4 M_{\odot})$  	& $0.5$     & $85.4218$       & $38.0655$ & 6.64    \\ \cline{2-5}
													\hline						    
												\end{tabular}
											\end{center}
										\end{table*}
										
										\section{Conclusions}
										In this article, we have analysed a new class of compact objects with pressure anisotropy in Heintzmann geometry \cite{Heintz}. We have considered $g_{tt}$ metric component as proposed by Heintzmann \cite{Heintz} and solving the EFEs, the $g_{rr}$ metric component is determined in presence of pressure anisotropy. Using the $g_{rr}$ and $g_{tt}$ metric components, we have developed a model for a class of compact objects. We have determined the maximum mass by maximising the radial sound velocity ($ v_{r}^2 = 1$) inside the star for arbitrary choice of stellar radius. It is noted that maximum masses and radii lie within the range $1.87-3.04~ M_{\odot}$  and $8-13$ Km in isotropic case. Maximum compactness and surface red-shift are found to be $0.3443$ and  $0.7919$ respectively.
										It is also noted that maximum mass of the compact objects increases with anisotropy $\alpha$ and tabulated in Table \ref{Tab1}. For $\alpha=0.5$ maximum masses lie within the range $1.99-3.23~ M_{\odot}$ and maximum compactness, surface red-shift are $0.3667$ and  $0.9367$ respectively for radius $(b)=8, 9, 10, 11, 13$ Km. It is evident from Table \ref{Tab1} that surface red-shift does not depend on the radius of the star but depends on the anisotropy parameter $\alpha$. 
										The physical parameters such as energy density ($\rho$), radial pressure ($p_{r}$), transverse pressure ($p_{t}$) and their gradients and pressure anisotropy ($\Delta$) are shown in Figs. \ref{density} - \ref{delta} respectively. Fulfilment of causality conditions are shown in Figs. \ref{figvr} - \ref{figvt}. Variation of surface red-shift is shown in Fig. \ref{figz} and it is evident that the maximum value of surface red-shift always lies below the value $(Z_{s})_{max} \leq 5.211 $ as predicted by Ivanov \cite{Ivanov}. 
										
										After the evaluation of maximum mass for chosen radius, we have determined the value of $\rho$ and $p_{r}$ at all interior points and using the values of $\rho$ and $p_{r}$, the EoSs are plotted. Using the method of curve fitting two possible EoSs of interior matter are obtained, one is linear and the other is quadratic, for arbitrary choice of radius and corresponding maximum mass and are tabulated in Table \ref{Tab5}. Fitted EoSs are shown in Fig.~\ref{figeospolb11}. We note that best fitted EoS in our model is quadratic. Linear and quadratic fitted EoS are then used to solve TOV equation to verify the value of maximum mass and radius. Mass-radius curve are shown in Figs.~\ref{figmrtov} and ~\ref{figmrtovq} for linear and quadratic fitted EoS respectively. We note some interesting results which are tabulated in Table \ref{Tab4}. The maximum mass obtained from geometrical consideration by maximising sound velocity is nearly equal to the maximum mass obtained from the solution of TOV equation using quadratic fitted EoS for the parameter space used here. Also the maximum radius from the solution of TOV equation approaches the arbitrary chosen value for higher anisotropy parameter. We have also shown the variation of mass with central density in Figs.~\ref{figmvdtov} and \ref{figmvdtovq} for linear and quadratic EoS respectively. From Figs.~\ref{figmvdtov} and ~\ref{figmvdtovq}, it is also noted that our model satisfies the Harrison-Zel'dovich-Novikov criterion \cite{Harrison,Zeldovich}. Maximum allowable central densities are $2.95504 \times 10^{15}~gm/cm^3$ and  $1.97003 \times 10^{15}~gm/cm^3$ corresponding to maximum masses for radii 9 and 11 Km respectively with $\alpha=0$ and quadratic EoS in this model. It is noted that the maximum value of central density depends on anisotropy parameter ($\alpha$) as tabulated in Table \ref{Tab5} for linear and quadratic EoS. From Figs. \ref{figsec} - \ref{figdec} it is evident that our proposed model satisfies all the necessary energy conditions.  Stability conditions namely TOV equation, Herrera cracking condition and Adiabatic index are also studied and are shown in Figs.  \ref{figtov} - \ref{figadia} respectively and it is noted that our model satisfies all stability criteria. We have determined the tidal love number and tidal de-formability for three different compact stars GW 170817, PSR J1614-2230 and PSR J0952-0607 and are tabulated in Table \ref{Tab3}. It is noted that the value of tidal love number and tidal de-formability decreases with increase of anisotropy. We have also determined the values of different parameters ( $a,~ A,~\Lambda$ ) to predict the estimated masses and radii of few recently observed compact objects such as PSR J1614-2230, PSR J0952-0607, GW 190814 and GW 170817 and are tabulated in Table \ref{Tab2}. It is noted from Table \ref{Tab2} that radius of a compact object depends on the value of $\alpha$. Higher is the value of $\alpha$, lower is the radius. The predicted radius of PSR J1614-2230 is found to be 10.30 Km for $\alpha=0$ which is same as proposed by Kalam et al. \cite{kalam} previously. The radius of companion object in GW 190814 event has been predicted by Maurya et al. \cite{maurya2} very recently having value $11.76^{+0.14}_{-0.19}$ Km in the context of gravitational decoupling in Einstein-Gauss-Bonnet gravity. Our model also predicts the value of radius within the range $11.56-11.90$ Km for $\alpha = 0.67 - 0.82$. Therefore, It is concluded that our model may predict a wide range of masses and radii of compact objects following the fulfilment of causality conditions, energy condition and stability criteria.
										
										\section{Acknowledgements}
										BD and KBG are thankful to CSIR for providing fellowship vide no: 09/1219 (0005)/2019 EMR-I and 09/1219 (0004)/2019 EMR-I respectively.
										
\end{document}